\documentclass[aps,pra,twocolumn, groupedaddress, floatfix, longbibliography]{revtex4-2}

\usepackage{amsmath,amssymb,amsfonts,amsthm,bm}
\usepackage{graphicx}
\usepackage{float}
\usepackage[english]{babel}
\usepackage{dsfont}
\usepackage{epstopdf}
\usepackage{soul}
\usepackage{color}
\usepackage{braket}
\usepackage[normalem]{ulem}
\usepackage{subfiles} 

\usepackage[colorlinks=true,
            linkcolor=blue,
            urlcolor=blue,
            citecolor=blue]{hyperref}

\definecolor{SMblue}{rgb}{0.3, 0.2, 1}

\definecolor{GRgreen}{rgb}{0.31,0.74,0.47}

\emergencystretch=\maxdimen
\usepackage{titlesec}
\titlespacing\subsection{0pt}{12pt plus 4pt minus 2pt}{12pt plus 2pt minus 2pt}
\titlespacing\section{0pt}{12pt plus 4pt minus 2pt}{7pt plus 2pt minus 2pt}
\titlespacing\subsection{0pt}{12pt plus 4pt minus 2pt}{7pt plus 2pt minus 2pt}

\AtBeginDocument{%
    \newwrite\bibnotes
    \def\bibnotesext{Notes.bib}
    \immediate\openout\bibnotes=\jobname\bibnotesext
    \immediate\write\bibnotes{@CONTROL{REVTEX41Control}}
    \immediate\write\bibnotes{@CONTROL{%
    apsrev41Control,author="08",editor="1",pages="1",title="0",year="1"}}
     \if@filesw
     \immediate\write\@auxout{\string\citation{apsrev41Control}}%
    \fi
}%

\begin{document}
\title{Nonlinear Switch and Spatial Lattice Solitons of Photonic ${\bf{ s}}$-${\bf{p}}$ Orbitals \vspace{-2mm}}

\author{Gayathry~Rajeevan}
\affiliation{\vspace{2mm} Department of Physics, Indian Institute of Science, Bangalore 560012, India}
\author{Sebabrata~Mukherjee }
\email{mukherjee@iisc.ac.in}
\affiliation{\vspace{2mm} Department of Physics, Indian Institute of Science, Bangalore 560012, India}

\begin{abstract}
We develop fs laser-fabricated asymmetric  couplers and zig-zag arrays consisting of single and two-mode waveguides with bipartite nonlinearity. The fundamental mode ($s$ orbital) is near resonance with the neighboring higher-order $p$ orbital, causing efficient light transfer at low power. Due to Kerr nonlinearity, the coupler works as an all-optical switch between $s$ and $p$ orbitals. Single- and double-peak spatial solitons of $s$-$p$ orbitals form in the lattice due to the bipartite nature of the on-site nonlinearity. We probe highly localized bulk and edge solitons, peaked at the $s$ and $p$ orbitals, spectrally residing in the photonic band gap. Our work will be important for exploring inter-orbital couplings and nonlinear interactions in intricate photonic devices. 
\vspace{-1mm}
\end{abstract}
\maketitle
Engineered waveguide arrays are a versatile platform for exploring intriguing transport phenomena ranging from unidirectional topological edge states to traveling solitons~\cite{garanovich2012light, ozawa2019topological, smirnova2020nonlinear, lederer2008discrete}. Most of the research in this field has been performed using waveguides supporting the fundamental mode, analogously the $s$ orbital. The exploration of higher orbital physics in discrete lattices is of great interest in the context of fundamental~\cite{li2016physics, cantillano2018observation} as well as applied science~\cite{
sorin1986highly, love2012mode,
birks2015photonic, gross2014three, guzman2021experimental}. Inter-orbital couplings can generate synthetic magnetic flux~\cite{jorg2020artificial, schulz2022photonic, caceres2022controlled}, paving a new route for creating photonic topological materials. Photonic orbitals can also act as a synthetic dimension~\cite{lustig2019photonic} 
and give rise to various emergent phenomena~\cite{krupa2017spatial, wu2019thermodynamic} in the presence of optical nonlinearity.

In this work, we consider a photonic lattice of $s$ 
and $p_y$ (henceforth mentioned as $p$)
orbitals -- a periodic array of single and
two-mode waveguides, where energy exchange happens among the $s$ orbital of the single mode waveguide and the $p$ orbital of neighboring waveguides.
Using the building block of the lattice, i.e., a two-waveguide $s$-$p$ coupler, we demonstrate an all-optical nonlinear switch, where the tunneling of light from $s$ to $p$ orbital is tuned by input power. %
In the array of $s$-$p$ orbitals, we demonstrate novel lattice solitons -- shape-preserving nonlinear states~\cite{christodoulides1988discrete, eisenberg1998discrete, szameit2006two} -- which appear when the linear diffraction of light is balanced by optical nonlinearity. Because of the bipartite nature of the nonlinear strength in the $s$-$p$ orbitals, we observe single- and double-peak solitons that are not found in traditional waveguide arrays.

\begin{figure}[t!]
\includegraphics[width=1\linewidth]{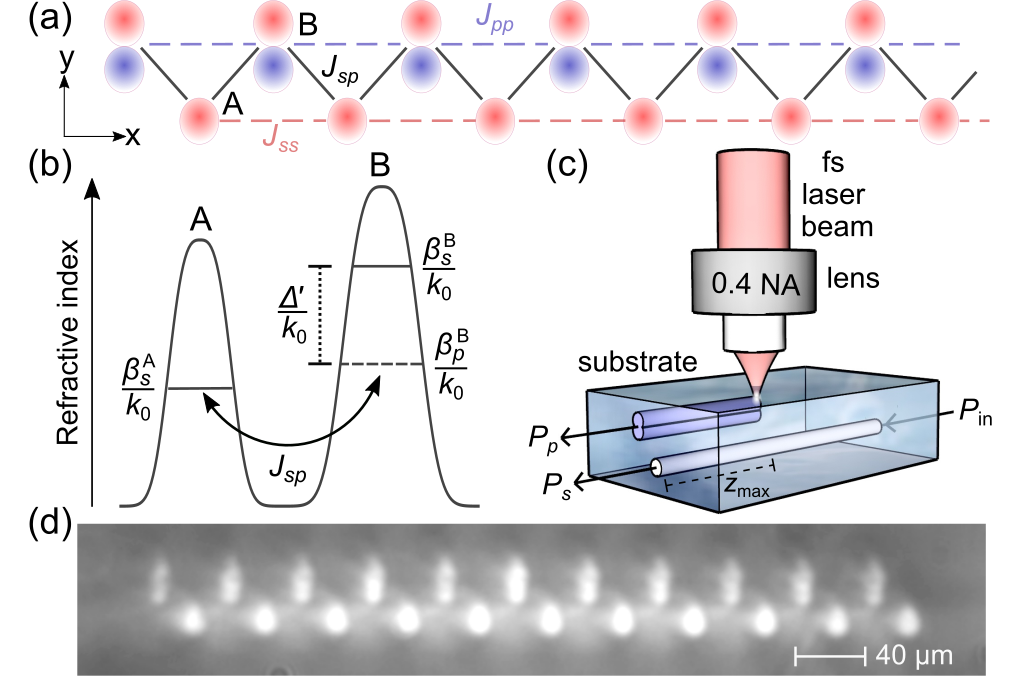}
    \caption{(a) Schematic of an $s$-$p$ orbital array with nearest ($J_{sp}$)
    and next-nearest neighbor couplings ($J_{ss}$ and $J_{pp}$).
     (b) A simplified sketch showing the refractive index profile and the modal refractive indices of the A and B sites of the lattice. The $s$ orbital of A site is near resonance with the $p$ orbital of B site, i.e., $\beta_s^{\text{A}}\approx \beta_p^{\text{B}}$, and $\Delta'\!\equiv\!\beta_s^{\text{B}}-\beta_p^{\text{B}}$ is designed to be $\gg \! J_{sp}$. $k_0$ is the free-space wave vector. %
    (c) Femtosecond laser writing of an $s$-$p$ coupler. Here, $z_{\text{max}}$ is the interaction length of the coupler, $P_{\text{in}}$, $P_s$, and $P_p$ are the input power, output power at the $s$ orbital and $p$ orbital, respectively.
    (d) Output facet image (cross-section) of a fs laser-written photonic $s$-$p$ orbital array with $22$ sites.}
    \label{Fig_model}
\end{figure}

{\it Model. $-$} Consider a one-dimensional zig-zag photonic lattice consisting of A and B sites per unit cell, as shown in Fig.~\ref{Fig_model}(a). The A site supports only the fundamental mode  of propagation constant $\beta_s^{\text{A}}$, and the refractive index profile of the B site is engineered such that the higher orbital $p$ is near resonance (phase matched) with the $s$ orbital of A site, i.e., $\beta_s^{\text{A}}\approx\beta_p^{\text{B}}$. The inter-site nearest-neighbor couplings between the $s$ and $p$ orbitals are denoted by $J_{sp}$ and next nearest-neighbor couplings among $s$-$s$ and $p$-$p$ orbitals are denoted by $J_{ss}$ and $J_{pp}$, respectively. When the $s$ orbital of A site is initially excited, light can only tunnel to the $p$ orbitals of the neighboring sites and vice versa. In other words, $(\beta_s^{\text{B}}-\beta_p^{\text{B}})/J_{sp}\! \gg \!1$, and hence, we shall now consider the dynamics of optical fields in the $s$ and $p$ orbitals of A and B sites, respectively.
In the scalar-paraxial approximation, the propagation of light in the $s$-$p$ orbital array 
can be governed by the following discrete nonlinear Schr{\"o}dinger equation~\cite{Suppmat} 
\begin{align}
 i\frac{\partial}{\partial z}a_j^s(z) \!=\! & -J_{sp}\big(b_j^p+ b_{j-1}^p \big) - J_{ss} \big(a_{j+1}^s + a_{j-1}^s \big) \nonumber\\
&-\beta_s^{\text{A}} a_{j}^s-g_s^{\text{A}}|a_j^s|^2 a_j^s \, , \label{DNLSE_sp1}\\
i\frac{\partial}{\partial z}b_j^p(z) \!=\! & -J_{sp} \big( a_j^s+ a_{j+1}^s \big)
-J_{pp} \big( b_{j+1}^p+b_{j-1}^p \big) \nonumber \\
&-\beta_p^{\text{B}} b_{j}^p-g_p^{\text{B}}|b_j^p|^2b_j^p\, ,
\label{DNLSE_sp}     
\end{align}
where $j$ indicates the index of the unit cell, $z$ denotes the propagation distance and $a_j^{s}$ ($b_j^{p}$) is proportional to the electric field envelope 
of the $s$ ($p$) 
orbital at the $j$-th unit cell. 
The nonlinearity in the system emerges due to self-focusing optical Kerr 
effect which is negligible at low optical power. The nonlinear strength is given by $g\!=\!2\pi n_2/(\lambda A_\text{eff})$,
where $n_2$ is the nonlinear refractive index, $\lambda$ is the wavelength of incident light and $A_\text{eff}$ is the effective area of the waveguide mode. %
Light at the $p$ orbital experiences a lower nonlinearity ($g_p^{\text{B}}\!<\!g_s^{\text{A}}$) owing to its larger effective area as well as the lower nonlinear refractive index of the B site.
%

\begin{figure}[t!]
    \includegraphics[width=1\linewidth]{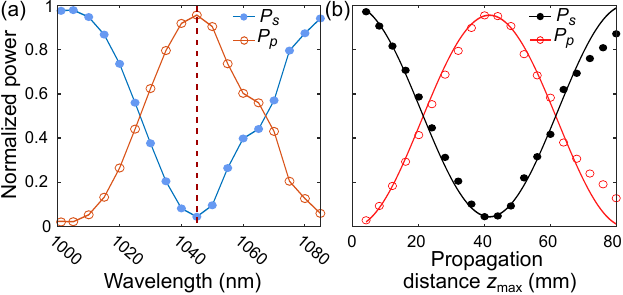}
    \caption{Linear characterization of $s$-$p$ couplers. (a) Measured variation of the normalized output power in the $s$ and $p$ orbitals ($P_s$ and $P_p$) as a function of wavelength of light for a $40$-mm-long $s$-$p$ coupler. Nearly full transfer ($ 96\%$) of power is observed at $\lambda\!=\!1045$~nm. (b) Normalized power $P_s$ and $P_p$ as a function of propagation distance at $\lambda\!=\!1045$~nm.}
    \label{Fig_coupler_lin}
\end{figure}

{\it Fabrication.$-$}
Waveguides were fabricated using fs laser writing~\cite{davis1996writing} in borosilicate (BK7) glass substrate. 
Circularly polarized $260$~fs (FWHM) laser pulses at $1030$~nm wavelength, $380$~nJ energy, and $500$~kHz repetition rate were focused inside an 80-mm-long 
substrate mounted on high-precision $x$-$y$-$z$ translation stages (Aerotech); Fig.~\ref{Fig_model}(c). Each single mode waveguide was created by translating the substrate twice through the focus of the laser beam at $4$~mm/s speed. To fabricate the multi-mode waveguides, a similar two-scan process was used with a vertical scan-to-scan separation of $6\,\mu$m. The optimum translation speeds of the lower and upper scans were found to be $1$~mm/s and $2$~mm/s, respectively, to obtain $\beta_s^{\text{A}}\approx\beta_p^{\text{B}}$. A transmission micrograph of the output facet of the zig-zag lattice is shown in Fig.~\ref{Fig_model}(d).

\begin{figure}[]
    \includegraphics[width=1\linewidth]{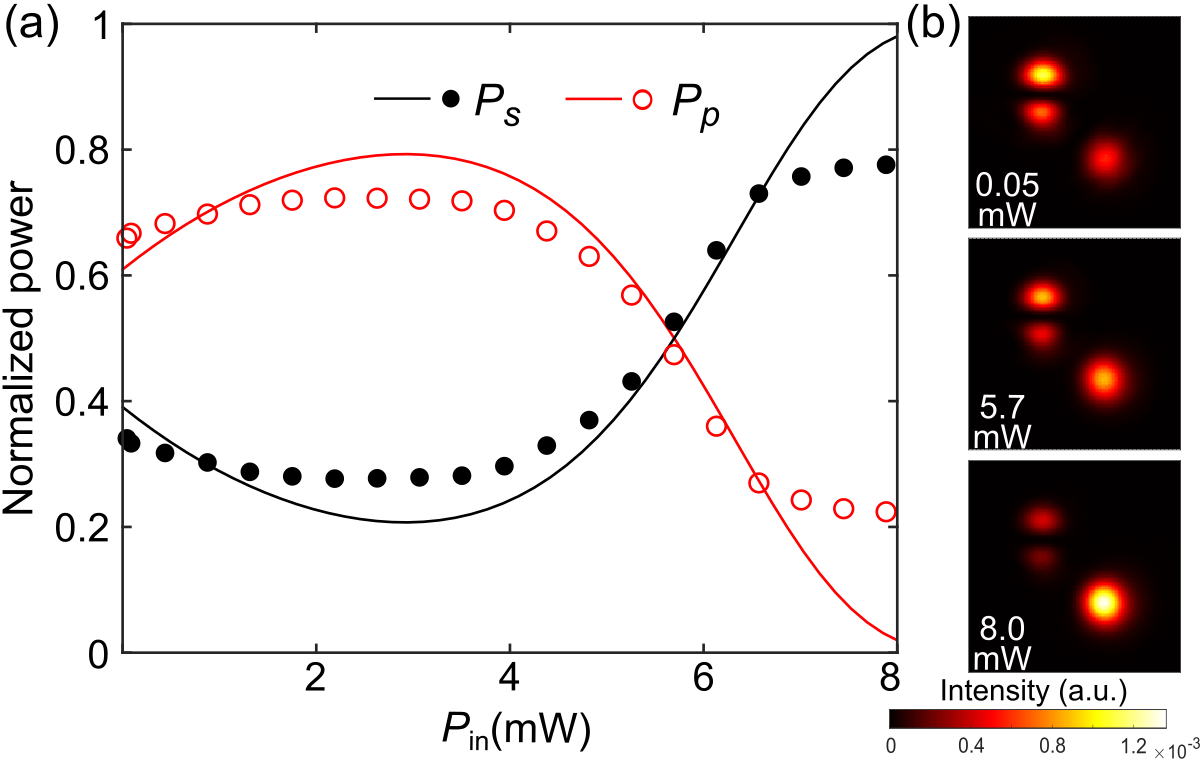}
    \caption{All-optical nonlinear switch of $s$-$p$ orbitals. Normalized power $P_s$ and $P_p$ as a function input power $P_{\text{in}}$ at $1030$~nm wavelength for a coupler of $z_{\text{max}}\!=\!36$~mm interaction length. The solid lines are obtained numerically. (b) Output intensity distributions at three different input powers. 
    }
    \label{Fig_coupler_NL}
\end{figure}

  \begin{figure}[b!]
    \includegraphics[width=1\linewidth]{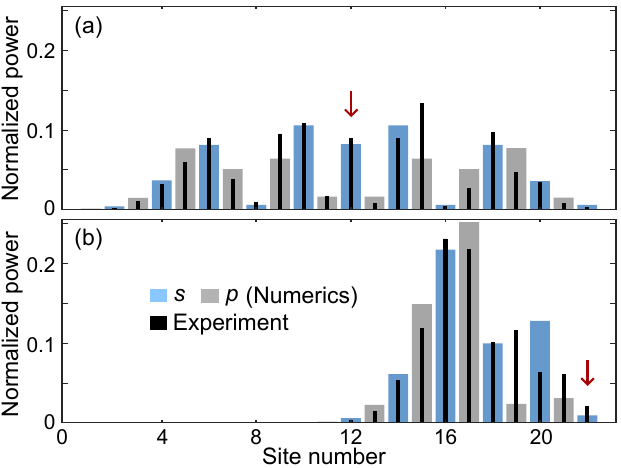}
    \caption{Linear characterization of the $s$-$p$ orbital array at $1030$~nm wavelength. (a) Distribution of optical power at the output of an $80$-mm-long array. Low-power light is initially coupled at site $12$, indicated by the red arrow. Numerically obtained powers at the $s$ and $p$ orbitals are indicated by blue and grey colors, respectively. Experimentally measured power values are shown in black. (b) Same as (a) with input excitation at the edge site, i.e., site $22$ (see also Fig.~\ref{Fig_model}(d)). 
    }
    \label{Fig_lattice_lin}
\end{figure}
{\it All-optical $s$-$p$ nonlinear switch.$-$} 
The building block of our zig-zag array is an $s$-$p$ coupler, consisting of two straight waveguides (A and B sites) -- see Ref.~\cite{eaton2006telecom} for traditional $s$-$s$ couplers. We fabricate twenty sets of $s$-$p$ couplers, each with an inter-waveguide spacing of $30$~$\mu$m and waveguide-to-waveguide angle of 45$^{\circ}$ relative to the vertical axis. As indicated in Fig.~\ref{Fig_model}(c), the length of the single mode A site was $80$~mm (fixed) for all couplers, and 
the interaction length was varied by changing the length of the B site
from $4$~mm to $80$~mm 
in steps of $4$~mm. These couplers were characterized by launching horizontally polarized light at the A site and imaging the output intensity pattern on a CMOS camera. For the linear experiments, a commercially available wavelength tunable super-continuum source (NKT Photonics) was used. 
Fig.~\ref{Fig_coupler_lin}(a) 
shows the measured normalized output powers ($P_s$ and $P_p$) for a $40$-mm-long $s$-$p$ coupler 
as a function of the wavelength of light. 
Nearly full transfer ($96\%$) of power is observed at $\lambda\!=\!1045$ nm, indicating that the $s$-orbital of A site is at near resonance with the $p$ site of B site at 
this wavelength. For further confirmation, we measured the
variation of 
$P_s$ and $P_p$
as a function of propagation distance at $\lambda\!=\!1045$~nm, see Fig.~\ref{Fig_coupler_lin}(b). 
By fitting the experimental data in Fig.~\ref{Fig_coupler_lin}(b) with the coupled-mode equations Eqs.~(\ref{DNLSE_sp1}, ~\ref{DNLSE_sp}) (considering a unit cell only) for a linear $s$-$p$ coupler,
$\Delta\! \equiv \!|\beta_s^{\text{A}}-\beta_p^{\text{B}}|$ and $J_{sp}$ were found to be $0.016$~mm$^{-1}$ and $0.037$~mm$^{-1}$,
respectively. Both $\Delta$ and $J_{sp}$ were found to vary with the wavelength of light -- in the supplementary material~\cite{Suppmat}, we present the variation of $P_s$, $P_p$ with $z_\text{max}$ for the same coupler at $1030$~nm wavelength.

To minimize wavelength broadening due to self-phase modulation, we used 
 $1.1$~ps down-chirped laser pulses 
 at $5$~kHz repetition rate and $1030$~nm wavelength for all nonlinear experiments. 
Additionally, the effect of nonlinear absorption and chromatic dispersion were found to be negligible. The propagation loss for a 
straight single mode waveguide was measured to be $\alpha_s\!=\!0.25$~dB/cm. 
In numerics, the effect of propagation loss in $s$ and $p$ orbitals are incorporated by adding 
$-i \frac{\alpha_s}{2} a_j^s$ and $-i \frac{\alpha_p}{2}b_j^p$
to the 
right-hand side of Eqs.~(\ref{DNLSE_sp1},~\ref{DNLSE_sp}), respectively.
We have assumed $\alpha_p\!=\!\alpha_s$ because the $s$-$p$ couplers exhibited similar transmission as that of an isolated single mode waveguide~\cite{Suppmat}. %
Fig.~\ref{Fig_coupler_NL}(a) shows the measured variation of normalized output power $P_s$ and $P_p$ of an $s$-$p$ coupler as a function of average input power $P_{\text{in}}$. Similar to the previous linear experiments, laser pulses are coupled to the $s$ orbital of the A waveguide. At low input power, $67\%$ of optical power is transferred to the nearest $p$ orbital of the B site after a propagation of $36$~mm. At a higher input power, due to Kerr nonlinearity, this power transfer is largely inhibited, and $78\%$ of light remains in the $s$ orbital.
Thus, the $s$-$p$ 
coupler acts as an all-optical switch between $s$ and $p$ orbitals.
We estimated $A_\text{eff}^p/A_\text{eff}^s$ and $n_2^{\text B}/ n_2^{\text A}$ for the device to be $1.7$ and $0.9$, respectively. The solid lines in Fig.~\ref{Fig_coupler_NL}(a) are obtained by solving Eqs.~(\ref{DNLSE_sp1}, \ref{DNLSE_sp}) for the nonlinear $s$-$p$ coupler. The fitted value of the nonlinear refractive index $n_2^{\text{A}}$ is $1.1 \times 10^{-20}$~m$^2$/W. Note that the values of $n_2$ for the waveguides are smaller than the pristine glass~\cite{flom2015ultrafast} due to laser writing process~\cite{blomer2006nonlinear, demetriou2017nonlinear}. 
\begin{figure}[t!]
\includegraphics[width=1\linewidth]{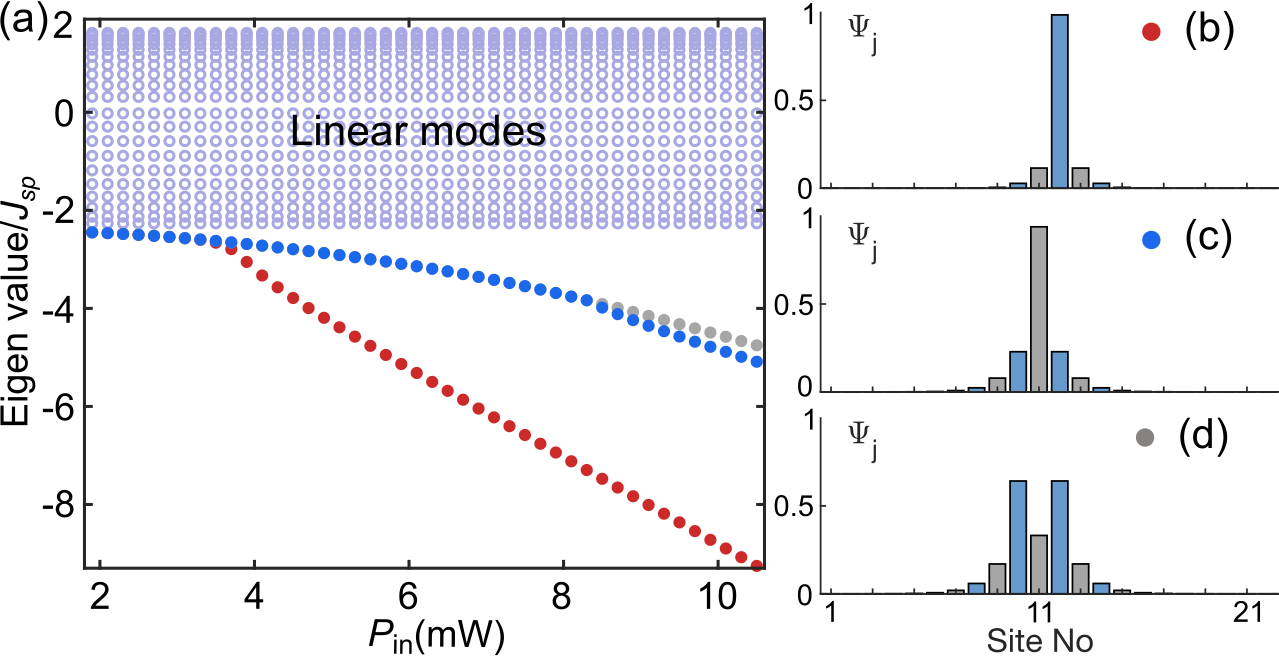}
    \caption{Nonlinear spectrum of the $s$-$p$ array with bipartite nonlinearity $g_s^\text{A}/g_p^\text{B}\!=\!1.87$. Three branches of solitons are shown by red, blue, and gray filled circles.
(b-d) Soliton wave functions of the three branches in (a) calculated at $P_{\text{in}}\!=\!10.5$~mW.
}
    \label{Fig_NL_spectrum}
\end{figure}

\begin{figure}[t!]
\includegraphics[width=1\linewidth]{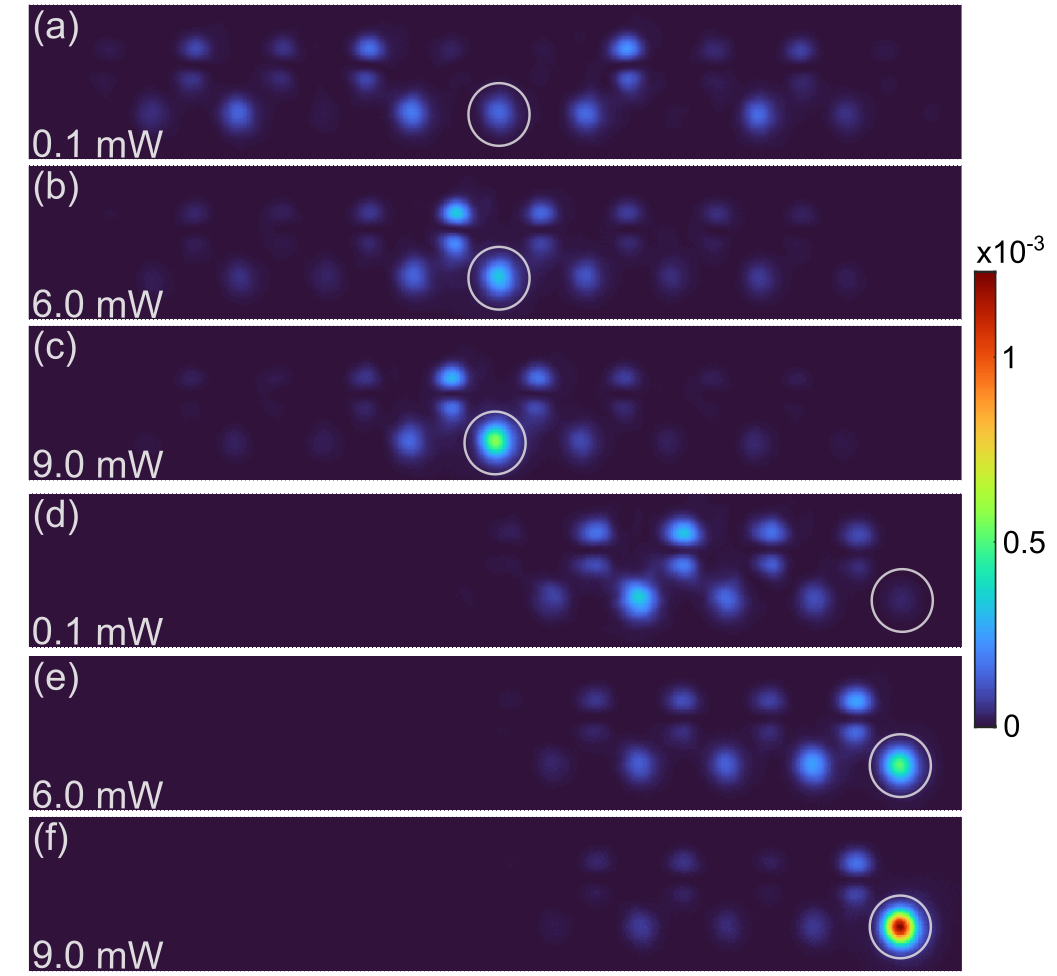}
    \caption{
    Measured intensity distributions at the output of an $80$-mm-long $s$-$p$ orbital array as a function of average input power indicated on each image. The white circle indicates where the light was coupled at the input. (a-c) and (d-f) show the formation of highly localized bulk and edge solitons, respectively, peaked at the A sites. 
    Each image is normalized such that the total power is $1$. }
    \label{Fig_soliton}
\end{figure}

{\it Solitons in $s$-$p$ orbital array.$-$} 
To study light transport in the $s$-$p$ array, we created an $80$-mm-long finite lattice of 22 sites with an inter-site spacing of $28\, \mu$m. The linear discrete diffraction is presented in Figs.~\ref{Fig_lattice_lin}(a, b) for input excitation at the central and edge $s$ sites, respectively, see also Figs.~\ref{Fig_soliton}(a, e). The experimental data agrees well with the numerics -- here $J_{sp}$, $J_{ss}\!\approx\!J_{pp}$, and $\Delta$ were estimated to be $0.045$, $0.006$ and $0.009$~mm$^{-1}$, respectively. The couplings decay exponentially with inter-waveguide spacing, hence, other long-range couplings are negligible. When we couple light into the $s$ orbital of the B site, no tunneling was observed, validating 
$\Delta'$ is significantly larger than $J_{sp}$.  

\begin{figure}[t!]
\includegraphics[width=1\linewidth]{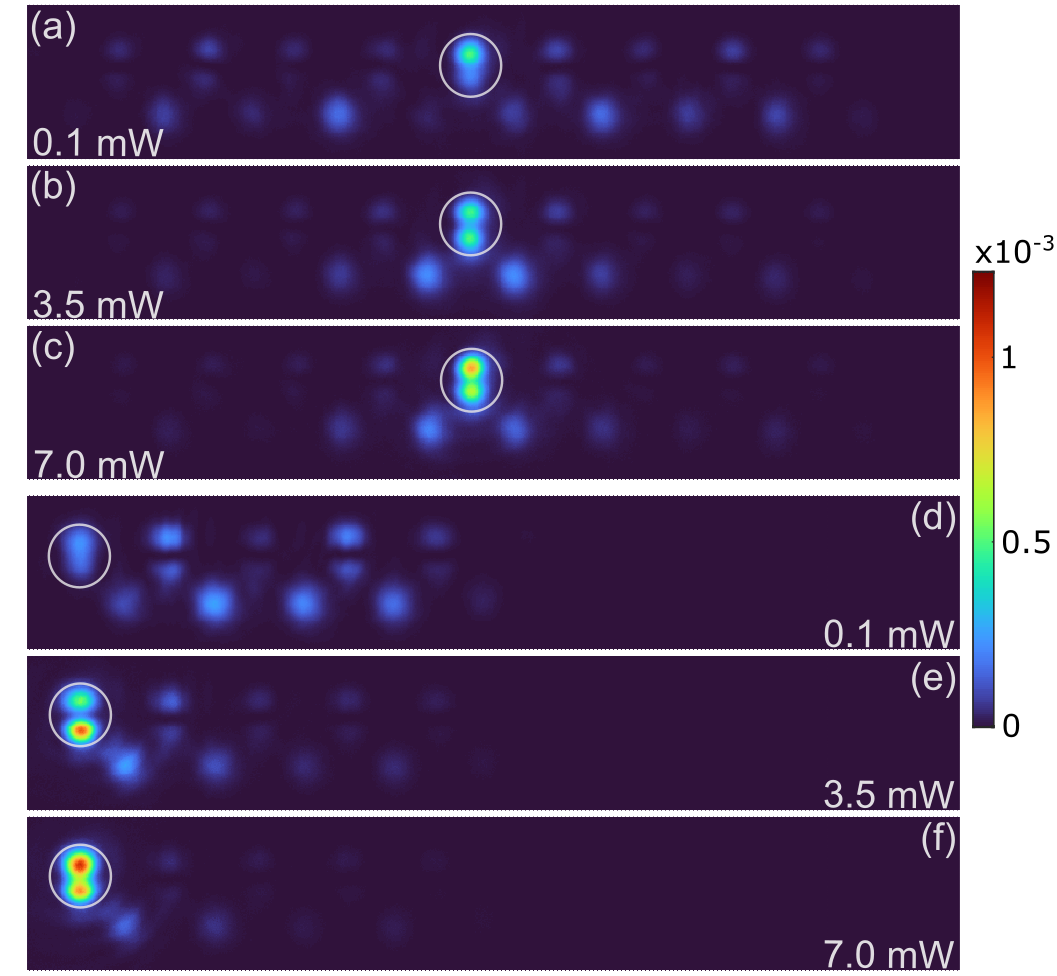}
    \caption{Similar to Fig.~\ref{Fig_soliton}. (a-c) and (d-f) show the formation of bulk and edge solitons, respectively, peaked at the $p$ orbital of B sites.}
    \label{Fig_P_soliton}
\end{figure}

We seek nonlinear soliton solutions in the $s$-$p$ array using self-consistency algorithm~\cite{cohen2003multiband}. 
Shape-preserving nonlinear solutions are iteratively calculated starting from initial guess solutions.
The nonlinear spectrum is presented in Fig.~\ref{Fig_NL_spectrum}(a), where three soliton families  are indicated by red, blue, and gray filled circles.
The red and blue colored solitons have a single peak at the $s$ orbital of A site and $p$ orbital of B site, respectively, see Figs.~\ref{Fig_NL_spectrum}(b,c). On the other hand, the gray-colored solitons centered at the B site have two peaks, Fig.~\ref{Fig_NL_spectrum}(d). 
Because of the bipartite nonlinear strength in the $s$-$p$ array, these soliton families are distinct from the typical solitons observed in the array of $s$ orbitals. Linear stability analysis~\cite{eilbeck1985discrete}  confirms that both types of single-peak solitons are stable at high power (10.5~mW), however, the double-peak one is linearly unstable; see supplementary material.
The single-peak solitons in Figs.~\ref{Fig_NL_spectrum}(b, c) can be experimentally probed by single-site excitation as shown below.

To experimentally study the nonlinear dynamics, we first launch light at the central and edge A sites, respectively, and measured output intensity patterns as a function of average input power, see Fig.~\ref{Fig_soliton}.
At low optical power, linear behavior is 
observed, and as the optical power is increased, the intensity pattern becomes increasingly localized, demonstrating highly localized spatial solitons peaked at the $s$ orbital of A site. %

Fig.~\ref{Fig_P_soliton} shows the output intensity distributions as a function of average input power launched at the central and edge B sites, respectively. 
The formation of highly localized bulk and edge solitons peaked at the $p$ orbital are clearly visible. 
%
In these experiments, optical power at the input is mostly ($75\pm5 \%$) coupled to the $p$ orbital of the B site -- the large coupling to the $p$ orbital is due to its asymmetric and bigger upper lob size.
The light coupled 
to the $s$ orbital of B site 
does not tunnel but contributes to the change in local nonlinear refractive index of the input site. 
As a result, nonlinearity-induced light localization in Fig.~\ref{Fig_P_soliton} is observed at a relatively lower input power compared to Fig.~\ref{Fig_soliton}.

In conclusion, we have studied linear and nonlinear light transport in fs laser-written mode-selective couplers and zig-zag arrays of $s$-$p$ orbitals.
The all-optical nonlinear switch of $s$-$p$ modes can have potential applications in on-chip mode-division multiplexing~\cite{richardson2013space, miller2013reconfigurable}. 
Due to the bipartite nature of the nonlinear strength, we observed 
novel families of bulk and edge solitons that can not be found in a traditional 1D photonic lattice.
Our work can be extended to explore light coupling and transport in other higher orbitals of few-mode waveguide systems.
Furthermore, inter-orbital couplings will be useful for predicting and realizing novel photonic topological systems~\cite{li2013topological}.


{\it Funding.$-$}
Start-up grant from the Indian Institute of Science (IISc), STARS (file no MoE-STARS/STARS-2/2023-0716) from the Ministry of Education, Govt. of India, and Start-up Research Grant (file no SRG/2022/002062)
from Science and Engineering Research Board (SERB).

{\it Acknowledgments.$-$}  
We acknowledge helpful discussions with Bhoomija Chaurasia, Archit Gajera, M.R. Shenoy, and Robert R.~Thomson. G.R.~thanks IISc for a PhD scholarship.


\begin{thebibliography}{31}%
\makeatletter
\providecommand \@ifxundefined [1]{%
 \@ifx{#1\undefined}
}%
\providecommand \@ifnum [1]{%
 \ifnum #1\expandafter \@firstoftwo
 \else \expandafter \@secondoftwo
 \fi
}%
\providecommand \@ifx [1]{%
 \ifx #1\expandafter \@firstoftwo
 \else \expandafter \@secondoftwo
 \fi
}%
\providecommand \natexlab [1]{#1}%
\providecommand \enquote  [1]{``#1''}%
\providecommand \bibnamefont  [1]{#1}%
\providecommand \bibfnamefont [1]{#1}%
\providecommand \citenamefont [1]{#1}%
\providecommand \href@noop [0]{\@secondoftwo}%
\providecommand \href [0]{\begingroup \@sanitize@url \@href}%
\providecommand \@href[1]{\@@startlink{#1}\@@href}%
\providecommand \@@href[1]{\endgroup#1\@@endlink}%
\providecommand \@sanitize@url [0]{\catcode `\\12\catcode `\$12\catcode `\&12\catcode `\#12\catcode `\^12\catcode `\_12\catcode `\%12\relax}%
\providecommand \@@startlink[1]{}%
\providecommand \@@endlink[0]{}%
\providecommand \url  [0]{\begingroup\@sanitize@url \@url }%
\providecommand \@url [1]{\endgroup\@href {#1}{\urlprefix }}%
\providecommand \urlprefix  [0]{URL }%
\providecommand \Eprint [0]{\href }%
\providecommand \doibase [0]{https://doi.org/}%
\providecommand \selectlanguage [0]{\@gobble}%
\providecommand \bibinfo  [0]{\@secondoftwo}%
\providecommand \bibfield  [0]{\@secondoftwo}%
\providecommand \translation [1]{[#1]}%
\providecommand \BibitemOpen [0]{}%
\providecommand \bibitemStop [0]{}%
\providecommand \bibitemNoStop [0]{.\EOS\space}%
\providecommand \EOS [0]{\spacefactor3000\relax}%
\providecommand \BibitemShut  [1]{\csname bibitem#1\endcsname}%
\let\auto@bib@innerbib\@empty
\bibitem [{\citenamefont {Garanovich}\ \emph {et~al.}(2012)\citenamefont {Garanovich}, \citenamefont {Longhi}, \citenamefont {Sukhorukov},\ and\ \citenamefont {Kivshar}}]{garanovich2012light}%
  \BibitemOpen
  \bibfield  {author} {\bibinfo {author} {\bibfnamefont {I.~L.}\ \bibnamefont {Garanovich}}, \bibinfo {author} {\bibfnamefont {S.}~\bibnamefont {Longhi}}, \bibinfo {author} {\bibfnamefont {A.~A.}\ \bibnamefont {Sukhorukov}},\ and\ \bibinfo {author} {\bibfnamefont {Y.~S.}\ \bibnamefont {Kivshar}},\ }\href {https://doi.org/https://doi.org/10.1016/j.physrep.2012.03.005} {\bibfield  {journal} {\bibinfo  {journal} {Phys. Rep.}\ }\textbf {\bibinfo {volume} {518}},\ \bibinfo {pages} {1} (\bibinfo {year} {2012})}\BibitemShut {NoStop}%
\bibitem [{\citenamefont {Ozawa}\ \emph {et~al.}(2019)\citenamefont {Ozawa}, \citenamefont {Price}, \citenamefont {Amo}, \citenamefont {Goldman}, \citenamefont {Hafezi}, \citenamefont {Lu}, \citenamefont {Rechtsman}, \citenamefont {Schuster}, \citenamefont {Simon}, \citenamefont {Zilberberg},\ and\ \citenamefont {Carusotto}}]{ozawa2019topological}%
  \BibitemOpen
  \bibfield  {author} {\bibinfo {author} {\bibfnamefont {T.}~\bibnamefont {Ozawa}}, \bibinfo {author} {\bibfnamefont {H.~M.}\ \bibnamefont {Price}}, \bibinfo {author} {\bibfnamefont {A.}~\bibnamefont {Amo}}, \bibinfo {author} {\bibfnamefont {N.}~\bibnamefont {Goldman}}, \bibinfo {author} {\bibfnamefont {M.}~\bibnamefont {Hafezi}}, \bibinfo {author} {\bibfnamefont {L.}~\bibnamefont {Lu}}, \bibinfo {author} {\bibfnamefont {M.~C.}\ \bibnamefont {Rechtsman}}, \bibinfo {author} {\bibfnamefont {D.}~\bibnamefont {Schuster}}, \bibinfo {author} {\bibfnamefont {J.}~\bibnamefont {Simon}}, \bibinfo {author} {\bibfnamefont {O.}~\bibnamefont {Zilberberg}},\ and\ \bibinfo {author} {\bibfnamefont {I.}~\bibnamefont {Carusotto}},\ }\href {https://doi.org/10.1103/RevModPhys.91.015006} {\bibfield  {journal} {\bibinfo  {journal} {Rev. Mod. Phys.}\ }\textbf {\bibinfo {volume} {91}},\ \bibinfo {pages} {015006} (\bibinfo {year} {2019})}\BibitemShut {NoStop}%
\bibitem [{\citenamefont {Smirnova}\ \emph {et~al.}(2020)\citenamefont {Smirnova}, \citenamefont {Leykam}, \citenamefont {Chong},\ and\ \citenamefont {Kivshar}}]{smirnova2020nonlinear}%
  \BibitemOpen
  \bibfield  {author} {\bibinfo {author} {\bibfnamefont {D.}~\bibnamefont {Smirnova}}, \bibinfo {author} {\bibfnamefont {D.}~\bibnamefont {Leykam}}, \bibinfo {author} {\bibfnamefont {Y.}~\bibnamefont {Chong}},\ and\ \bibinfo {author} {\bibfnamefont {Y.}~\bibnamefont {Kivshar}},\ }\bibfield  {journal} {\bibinfo  {journal} {Appl. Phys. Rev.}\ }\textbf {\bibinfo {volume} {7}},\ \href {https://doi.org/https://doi.org/10.1063/1.5142397} {https://doi.org/10.1063/1.5142397} (\bibinfo {year} {2020})\BibitemShut {NoStop}%
\bibitem [{\citenamefont {Lederer}\ \emph {et~al.}(2008)\citenamefont {Lederer}, \citenamefont {Stegeman}, \citenamefont {Christodoulides}, \citenamefont {Assanto}, \citenamefont {Segev},\ and\ \citenamefont {Silberberg}}]{lederer2008discrete}%
  \BibitemOpen
  \bibfield  {author} {\bibinfo {author} {\bibfnamefont {F.}~\bibnamefont {Lederer}}, \bibinfo {author} {\bibfnamefont {G.~I.}\ \bibnamefont {Stegeman}}, \bibinfo {author} {\bibfnamefont {D.~N.}\ \bibnamefont {Christodoulides}}, \bibinfo {author} {\bibfnamefont {G.}~\bibnamefont {Assanto}}, \bibinfo {author} {\bibfnamefont {M.}~\bibnamefont {Segev}},\ and\ \bibinfo {author} {\bibfnamefont {Y.}~\bibnamefont {Silberberg}},\ }\href {https://doi.org/https://doi.org/10.1016/j.physrep.2008.04.004} {\bibfield  {journal} {\bibinfo  {journal} {Phys. Rep.}\ }\textbf {\bibinfo {volume} {463}},\ \bibinfo {pages} {1} (\bibinfo {year} {2008})}\BibitemShut {NoStop}%
\bibitem [{\citenamefont {Li}\ and\ \citenamefont {Liu}(2016)}]{li2016physics}%
  \BibitemOpen
  \bibfield  {author} {\bibinfo {author} {\bibfnamefont {X.}~\bibnamefont {Li}}\ and\ \bibinfo {author} {\bibfnamefont {W.~V.}\ \bibnamefont {Liu}},\ }\href {https://doi.org/10.1088/0034-4885/79/11/116401} {\bibfield  {journal} {\bibinfo  {journal} {Rep. Prog. Phys.}\ }\textbf {\bibinfo {volume} {79}},\ \bibinfo {pages} {116401} (\bibinfo {year} {2016})}\BibitemShut {NoStop}%
\bibitem [{\citenamefont {Cantillano}\ \emph {et~al.}(2018)\citenamefont {Cantillano}, \citenamefont {Mukherjee}, \citenamefont {Morales-Inostroza}, \citenamefont {Real}, \citenamefont {C{\'a}ceres-Aravena}, \citenamefont {Hermann-Avigliano}, \citenamefont {Thomson},\ and\ \citenamefont {Vicencio}}]{cantillano2018observation}%
  \BibitemOpen
  \bibfield  {author} {\bibinfo {author} {\bibfnamefont {C.}~\bibnamefont {Cantillano}}, \bibinfo {author} {\bibfnamefont {S.}~\bibnamefont {Mukherjee}}, \bibinfo {author} {\bibfnamefont {L.}~\bibnamefont {Morales-Inostroza}}, \bibinfo {author} {\bibfnamefont {B.}~\bibnamefont {Real}}, \bibinfo {author} {\bibfnamefont {G.}~\bibnamefont {C{\'a}ceres-Aravena}}, \bibinfo {author} {\bibfnamefont {C.}~\bibnamefont {Hermann-Avigliano}}, \bibinfo {author} {\bibfnamefont {R.~R.}\ \bibnamefont {Thomson}},\ and\ \bibinfo {author} {\bibfnamefont {R.~A.}\ \bibnamefont {Vicencio}},\ }\href {https://doi.org/10.1088/1367-2630/aab483} {\bibfield  {journal} {\bibinfo  {journal} {New J. Phys.}\ }\textbf {\bibinfo {volume} {20}},\ \bibinfo {pages} {033028} (\bibinfo {year} {2018})}\BibitemShut {NoStop}%
\bibitem [{\citenamefont {Sorin}\ \emph {et~al.}(1986)\citenamefont {Sorin}, \citenamefont {Kim},\ and\ \citenamefont {Shaw}}]{sorin1986highly}%
  \BibitemOpen
  \bibfield  {author} {\bibinfo {author} {\bibfnamefont {W.~V.}\ \bibnamefont {Sorin}}, \bibinfo {author} {\bibfnamefont {B.~Y.}\ \bibnamefont {Kim}},\ and\ \bibinfo {author} {\bibfnamefont {H.~J.}\ \bibnamefont {Shaw}},\ }\href {https://doi.org/https://doi.org/10.1364/OL.11.000581} {\bibfield  {journal} {\bibinfo  {journal} {Opt. Lett.}\ }\textbf {\bibinfo {volume} {11}},\ \bibinfo {pages} {581} (\bibinfo {year} {1986})}\BibitemShut {NoStop}%
\bibitem [{\citenamefont {Love}\ and\ \citenamefont {Riesen}(2012)}]{love2012mode}%
  \BibitemOpen
  \bibfield  {author} {\bibinfo {author} {\bibfnamefont {J.~D.}\ \bibnamefont {Love}}\ and\ \bibinfo {author} {\bibfnamefont {N.}~\bibnamefont {Riesen}},\ }\href {https://doi.org/https://doi.org/10.1364/OL.37.003990} {\bibfield  {journal} {\bibinfo  {journal} {Opt. Lett.}\ }\textbf {\bibinfo {volume} {37}},\ \bibinfo {pages} {3990} (\bibinfo {year} {2012})}\BibitemShut {NoStop}%
\bibitem [{\citenamefont {Birks}\ \emph {et~al.}(2015)\citenamefont {Birks}, \citenamefont {Gris-S{\'a}nchez}, \citenamefont {Yerolatsitis}, \citenamefont {Leon-Saval},\ and\ \citenamefont {Thomson}}]{birks2015photonic}%
  \BibitemOpen
  \bibfield  {author} {\bibinfo {author} {\bibfnamefont {T.~A.}\ \bibnamefont {Birks}}, \bibinfo {author} {\bibfnamefont {I.}~\bibnamefont {Gris-S{\'a}nchez}}, \bibinfo {author} {\bibfnamefont {S.}~\bibnamefont {Yerolatsitis}}, \bibinfo {author} {\bibfnamefont {S.}~\bibnamefont {Leon-Saval}},\ and\ \bibinfo {author} {\bibfnamefont {R.~R.}\ \bibnamefont {Thomson}},\ }\href {https://doi.org/https://doi.org/10.1364/AOP.7.000107} {\bibfield  {journal} {\bibinfo  {journal} {Adv. Opt. Photon.}\ }\textbf {\bibinfo {volume} {7}},\ \bibinfo {pages} {107} (\bibinfo {year} {2015})}\BibitemShut {NoStop}%
\bibitem [{\citenamefont {Gross}\ \emph {et~al.}(2014)\citenamefont {Gross}, \citenamefont {Riesen}, \citenamefont {Love},\ and\ \citenamefont {Withford}}]{gross2014three}%
  \BibitemOpen
  \bibfield  {author} {\bibinfo {author} {\bibfnamefont {S.}~\bibnamefont {Gross}}, \bibinfo {author} {\bibfnamefont {N.}~\bibnamefont {Riesen}}, \bibinfo {author} {\bibfnamefont {J.~D.}\ \bibnamefont {Love}},\ and\ \bibinfo {author} {\bibfnamefont {M.~J.}\ \bibnamefont {Withford}},\ }\href {https://doi.org/https://doi.org/10.1002/lpor.201400078} {\bibfield  {journal} {\bibinfo  {journal} {Laser Photonics Rev.}\ }\textbf {\bibinfo {volume} {8}},\ \bibinfo {pages} {L81} (\bibinfo {year} {2014})}\BibitemShut {NoStop}%
\bibitem [{\citenamefont {Guzm{\'a}n-Silva}\ \emph {et~al.}(2021)\citenamefont {Guzm{\'a}n-Silva}, \citenamefont {C{\'a}ceres-Aravena},\ and\ \citenamefont {Vicencio}}]{guzman2021experimental}%
  \BibitemOpen
  \bibfield  {author} {\bibinfo {author} {\bibfnamefont {D.}~\bibnamefont {Guzm{\'a}n-Silva}}, \bibinfo {author} {\bibfnamefont {G.}~\bibnamefont {C{\'a}ceres-Aravena}},\ and\ \bibinfo {author} {\bibfnamefont {R.~A.}\ \bibnamefont {Vicencio}},\ }\href {https://doi.org/10.1103/PhysRevLett.127.066601} {\bibfield  {journal} {\bibinfo  {journal} {Phys. Rev. Lett.}\ }\textbf {\bibinfo {volume} {127}},\ \bibinfo {pages} {066601} (\bibinfo {year} {2021})}\BibitemShut {NoStop}%
\bibitem [{\citenamefont {J{\"o}rg}\ \emph {et~al.}(2020)\citenamefont {J{\"o}rg}, \citenamefont {Queralt{\'o}}, \citenamefont {Kremer}, \citenamefont {Pelegr{\'\i}}, \citenamefont {Schulz}, \citenamefont {Szameit}, \citenamefont {von Freymann}, \citenamefont {Mompart},\ and\ \citenamefont {Ahufinger}}]{jorg2020artificial}%
  \BibitemOpen
  \bibfield  {author} {\bibinfo {author} {\bibfnamefont {C.}~\bibnamefont {J{\"o}rg}}, \bibinfo {author} {\bibfnamefont {G.}~\bibnamefont {Queralt{\'o}}}, \bibinfo {author} {\bibfnamefont {M.}~\bibnamefont {Kremer}}, \bibinfo {author} {\bibfnamefont {G.}~\bibnamefont {Pelegr{\'\i}}}, \bibinfo {author} {\bibfnamefont {J.}~\bibnamefont {Schulz}}, \bibinfo {author} {\bibfnamefont {A.}~\bibnamefont {Szameit}}, \bibinfo {author} {\bibfnamefont {G.}~\bibnamefont {von Freymann}}, \bibinfo {author} {\bibfnamefont {J.}~\bibnamefont {Mompart}},\ and\ \bibinfo {author} {\bibfnamefont {V.}~\bibnamefont {Ahufinger}},\ }\href {https://doi.org/https://doi.org/10.1038/s41377-020-00385-6} {\bibfield  {journal} {\bibinfo  {journal} {Light Sci. Appl.}\ }\textbf {\bibinfo {volume} {9}},\ \bibinfo {pages} {150} (\bibinfo {year} {2020})}\BibitemShut {NoStop}%
\bibitem [{\citenamefont {Schulz}\ \emph {et~al.}(2022)\citenamefont {Schulz}, \citenamefont {Noh}, \citenamefont {Benalcazar}, \citenamefont {Bahl},\ and\ \citenamefont {von Freymann}}]{schulz2022photonic}%
  \BibitemOpen
  \bibfield  {author} {\bibinfo {author} {\bibfnamefont {J.}~\bibnamefont {Schulz}}, \bibinfo {author} {\bibfnamefont {J.}~\bibnamefont {Noh}}, \bibinfo {author} {\bibfnamefont {W.~A.}\ \bibnamefont {Benalcazar}}, \bibinfo {author} {\bibfnamefont {G.}~\bibnamefont {Bahl}},\ and\ \bibinfo {author} {\bibfnamefont {G.}~\bibnamefont {von Freymann}},\ }\href {https://doi.org/https://doi.org/10.1038/s41467-022-33894-6} {\bibfield  {journal} {\bibinfo  {journal} {Nat. Commun.}\ }\textbf {\bibinfo {volume} {13}},\ \bibinfo {pages} {6597} (\bibinfo {year} {2022})}\BibitemShut {NoStop}%
\bibitem [{\citenamefont {C{\'a}ceres-Aravena}\ \emph {et~al.}(2022)\citenamefont {C{\'a}ceres-Aravena}, \citenamefont {Guzm{\'a}n-Silva}, \citenamefont {Salinas},\ and\ \citenamefont {Vicencio}}]{caceres2022controlled}%
  \BibitemOpen
  \bibfield  {author} {\bibinfo {author} {\bibfnamefont {G.}~\bibnamefont {C{\'a}ceres-Aravena}}, \bibinfo {author} {\bibfnamefont {D.}~\bibnamefont {Guzm{\'a}n-Silva}}, \bibinfo {author} {\bibfnamefont {I.}~\bibnamefont {Salinas}},\ and\ \bibinfo {author} {\bibfnamefont {R.~A.}\ \bibnamefont {Vicencio}},\ }\href {https://doi.org/10.1103/PhysRevLett.128.256602} {\bibfield  {journal} {\bibinfo  {journal} {Phys. Rev. Lett.}\ }\textbf {\bibinfo {volume} {128}},\ \bibinfo {pages} {256602} (\bibinfo {year} {2022})}\BibitemShut {NoStop}%
\bibitem [{\citenamefont {Lustig}\ \emph {et~al.}(2019)\citenamefont {Lustig}, \citenamefont {Weimann}, \citenamefont {Plotnik}, \citenamefont {Lumer}, \citenamefont {Bandres}, \citenamefont {Szameit},\ and\ \citenamefont {Segev}}]{lustig2019photonic}%
  \BibitemOpen
  \bibfield  {author} {\bibinfo {author} {\bibfnamefont {E.}~\bibnamefont {Lustig}}, \bibinfo {author} {\bibfnamefont {S.}~\bibnamefont {Weimann}}, \bibinfo {author} {\bibfnamefont {Y.}~\bibnamefont {Plotnik}}, \bibinfo {author} {\bibfnamefont {Y.}~\bibnamefont {Lumer}}, \bibinfo {author} {\bibfnamefont {M.~A.}\ \bibnamefont {Bandres}}, \bibinfo {author} {\bibfnamefont {A.}~\bibnamefont {Szameit}},\ and\ \bibinfo {author} {\bibfnamefont {M.}~\bibnamefont {Segev}},\ }\href {https://doi.org/https://doi.org/10.1038/s41586-019-0943-7} {\bibfield  {journal} {\bibinfo  {journal} {Nature}\ }\textbf {\bibinfo {volume} {567}},\ \bibinfo {pages} {356} (\bibinfo {year} {2019})}\BibitemShut {NoStop}%
\bibitem [{\citenamefont {Krupa}\ \emph {et~al.}(2017)\citenamefont {Krupa}, \citenamefont {Tonello}, \citenamefont {Shalaby}, \citenamefont {Fabert}, \citenamefont {Barth{\'e}l{\'e}my}, \citenamefont {Millot}, \citenamefont {Wabnitz},\ and\ \citenamefont {Couderc}}]{krupa2017spatial}%
  \BibitemOpen
  \bibfield  {author} {\bibinfo {author} {\bibfnamefont {K.}~\bibnamefont {Krupa}}, \bibinfo {author} {\bibfnamefont {A.}~\bibnamefont {Tonello}}, \bibinfo {author} {\bibfnamefont {B.~M.}\ \bibnamefont {Shalaby}}, \bibinfo {author} {\bibfnamefont {M.}~\bibnamefont {Fabert}}, \bibinfo {author} {\bibfnamefont {A.}~\bibnamefont {Barth{\'e}l{\'e}my}}, \bibinfo {author} {\bibfnamefont {G.}~\bibnamefont {Millot}}, \bibinfo {author} {\bibfnamefont {S.}~\bibnamefont {Wabnitz}},\ and\ \bibinfo {author} {\bibfnamefont {V.}~\bibnamefont {Couderc}},\ }\href {https://doi.org/https://doi.org/10.1038/nphoton.2017.32} {\bibfield  {journal} {\bibinfo  {journal} {Nat. Photonics.}\ }\textbf {\bibinfo {volume} {11}},\ \bibinfo {pages} {237} (\bibinfo {year} {2017})}\BibitemShut {NoStop}%
\bibitem [{\citenamefont {Wu}\ \emph {et~al.}(2019)\citenamefont {Wu}, \citenamefont {Hassan},\ and\ \citenamefont {Christodoulides}}]{wu2019thermodynamic}%
  \BibitemOpen
  \bibfield  {author} {\bibinfo {author} {\bibfnamefont {F.~O.}\ \bibnamefont {Wu}}, \bibinfo {author} {\bibfnamefont {A.~U.}\ \bibnamefont {Hassan}},\ and\ \bibinfo {author} {\bibfnamefont {D.~N.}\ \bibnamefont {Christodoulides}},\ }\href {https://doi.org/https://doi.org/10.1038/s41566-019-0501-8} {\bibfield  {journal} {\bibinfo  {journal} {Nat. Photonics.}\ }\textbf {\bibinfo {volume} {13}},\ \bibinfo {pages} {776} (\bibinfo {year} {2019})}\BibitemShut {NoStop}%
\bibitem [{\citenamefont {Christodoulides}\ and\ \citenamefont {Joseph}(1988)}]{christodoulides1988discrete}%
  \BibitemOpen
  \bibfield  {author} {\bibinfo {author} {\bibfnamefont {D.}~\bibnamefont {Christodoulides}}\ and\ \bibinfo {author} {\bibfnamefont {R.}~\bibnamefont {Joseph}},\ }\href {https://doi.org/https://doi.org/10.1364/OL.13.000794} {\bibfield  {journal} {\bibinfo  {journal} {Opt. Lett.}\ }\textbf {\bibinfo {volume} {13}},\ \bibinfo {pages} {794} (\bibinfo {year} {1988})}\BibitemShut {NoStop}%
\bibitem [{\citenamefont {Eisenberg}\ \emph {et~al.}(1998)\citenamefont {Eisenberg}, \citenamefont {Silberberg}, \citenamefont {Morandotti}, \citenamefont {Boyd},\ and\ \citenamefont {Aitchison}}]{eisenberg1998discrete}%
  \BibitemOpen
  \bibfield  {author} {\bibinfo {author} {\bibfnamefont {H.~S.}\ \bibnamefont {Eisenberg}}, \bibinfo {author} {\bibfnamefont {Y.}~\bibnamefont {Silberberg}}, \bibinfo {author} {\bibfnamefont {R.}~\bibnamefont {Morandotti}}, \bibinfo {author} {\bibfnamefont {A.~R.}\ \bibnamefont {Boyd}},\ and\ \bibinfo {author} {\bibfnamefont {J.~S.}\ \bibnamefont {Aitchison}},\ }\href {https://doi.org/10.1103/PhysRevLett.81.3383} {\bibfield  {journal} {\bibinfo  {journal} {Phys. Rev. Lett.}\ }\textbf {\bibinfo {volume} {81}},\ \bibinfo {pages} {3383} (\bibinfo {year} {1998})}\BibitemShut {NoStop}%
\bibitem [{\citenamefont {Szameit}\ \emph {et~al.}(2006)\citenamefont {Szameit}, \citenamefont {Burghoff}, \citenamefont {Pertsch}, \citenamefont {Nolte}, \citenamefont {T{\"u}nnermann},\ and\ \citenamefont {Lederer}}]{szameit2006two}%
  \BibitemOpen
  \bibfield  {author} {\bibinfo {author} {\bibfnamefont {A.}~\bibnamefont {Szameit}}, \bibinfo {author} {\bibfnamefont {J.}~\bibnamefont {Burghoff}}, \bibinfo {author} {\bibfnamefont {T.}~\bibnamefont {Pertsch}}, \bibinfo {author} {\bibfnamefont {S.}~\bibnamefont {Nolte}}, \bibinfo {author} {\bibfnamefont {A.}~\bibnamefont {T{\"u}nnermann}},\ and\ \bibinfo {author} {\bibfnamefont {F.}~\bibnamefont {Lederer}},\ }\href {https://doi.org/https://doi.org/10.1364/OE.14.006055} {\bibfield  {journal} {\bibinfo  {journal} {Opt. Express}\ }\textbf {\bibinfo {volume} {14}},\ \bibinfo {pages} {6055} (\bibinfo {year} {2006})}\BibitemShut {NoStop}%
\bibitem [{Sup()}]{Suppmat}%
  \BibitemOpen
  \href@noop {} {}\bibinfo {note} {See supplementary materials}\BibitemShut {NoStop}%
\bibitem [{\citenamefont {Davis}\ \emph {et~al.}(1996)\citenamefont {Davis}, \citenamefont {Miura}, \citenamefont {Sugimoto},\ and\ \citenamefont {Hirao}}]{davis1996writing}%
  \BibitemOpen
  \bibfield  {author} {\bibinfo {author} {\bibfnamefont {K.~M.}\ \bibnamefont {Davis}}, \bibinfo {author} {\bibfnamefont {K.}~\bibnamefont {Miura}}, \bibinfo {author} {\bibfnamefont {N.}~\bibnamefont {Sugimoto}},\ and\ \bibinfo {author} {\bibfnamefont {K.}~\bibnamefont {Hirao}},\ }\href {https://doi.org/https://doi.org/10.1364/OL.21.001729} {\bibfield  {journal} {\bibinfo  {journal} {Opt. Lett.}\ }\textbf {\bibinfo {volume} {21}},\ \bibinfo {pages} {1729} (\bibinfo {year} {1996})}\BibitemShut {NoStop}%
\bibitem [{\citenamefont {Eaton}\ \emph {et~al.}(2006)\citenamefont {Eaton}, \citenamefont {Chen}, \citenamefont {Zhang}, \citenamefont {Zhang}, \citenamefont {Iyer}, \citenamefont {Aitchison},\ and\ \citenamefont {Herman}}]{eaton2006telecom}%
  \BibitemOpen
  \bibfield  {author} {\bibinfo {author} {\bibfnamefont {S.}~\bibnamefont {Eaton}}, \bibinfo {author} {\bibfnamefont {W.}~\bibnamefont {Chen}}, \bibinfo {author} {\bibfnamefont {L.}~\bibnamefont {Zhang}}, \bibinfo {author} {\bibfnamefont {H.}~\bibnamefont {Zhang}}, \bibinfo {author} {\bibfnamefont {R.}~\bibnamefont {Iyer}}, \bibinfo {author} {\bibfnamefont {J.}~\bibnamefont {Aitchison}},\ and\ \bibinfo {author} {\bibfnamefont {P.}~\bibnamefont {Herman}},\ }\href {https://doi.org/https://doi.org/10.1109/LPT.2006.884241} {\bibfield  {journal} {\bibinfo  {journal} {IEEE Photonics Technol. Lett.}\ }\textbf {\bibinfo {volume} {20}},\ \bibinfo {pages} {2174} (\bibinfo {year} {2006})}\BibitemShut {NoStop}%
\bibitem [{\citenamefont {Flom}\ \emph {et~al.}(2015)\citenamefont {Flom}, \citenamefont {Beadie}, \citenamefont {Bayya}, \citenamefont {Shaw},\ and\ \citenamefont {Auxier}}]{flom2015ultrafast}%
  \BibitemOpen
  \bibfield  {author} {\bibinfo {author} {\bibfnamefont {S.~R.}\ \bibnamefont {Flom}}, \bibinfo {author} {\bibfnamefont {G.}~\bibnamefont {Beadie}}, \bibinfo {author} {\bibfnamefont {S.~S.}\ \bibnamefont {Bayya}}, \bibinfo {author} {\bibfnamefont {B.}~\bibnamefont {Shaw}},\ and\ \bibinfo {author} {\bibfnamefont {J.~M.}\ \bibnamefont {Auxier}},\ }\href {https://doi.org/https://doi.org/10.1364/AO.54.00F123} {\bibfield  {journal} {\bibinfo  {journal} {Appl. Opt.}\ }\textbf {\bibinfo {volume} {54}},\ \bibinfo {pages} {F123} (\bibinfo {year} {2015})}\BibitemShut {NoStop}%
\bibitem [{\citenamefont {Bl{\"o}mer}\ \emph {et~al.}(2006)\citenamefont {Bl{\"o}mer}, \citenamefont {Szameit}, \citenamefont {Dreisow}, \citenamefont {Schreiber}, \citenamefont {Nolte},\ and\ \citenamefont {T{\"u}nnermann}}]{blomer2006nonlinear}%
  \BibitemOpen
  \bibfield  {author} {\bibinfo {author} {\bibfnamefont {D.}~\bibnamefont {Bl{\"o}mer}}, \bibinfo {author} {\bibfnamefont {A.}~\bibnamefont {Szameit}}, \bibinfo {author} {\bibfnamefont {F.}~\bibnamefont {Dreisow}}, \bibinfo {author} {\bibfnamefont {T.}~\bibnamefont {Schreiber}}, \bibinfo {author} {\bibfnamefont {S.}~\bibnamefont {Nolte}},\ and\ \bibinfo {author} {\bibfnamefont {A.}~\bibnamefont {T{\"u}nnermann}},\ }\href {https://doi.org/https://doi.org/10.1364/OE.14.002151} {\bibfield  {journal} {\bibinfo  {journal} {Opt. Express}\ }\textbf {\bibinfo {volume} {14}},\ \bibinfo {pages} {2151} (\bibinfo {year} {2006})}\BibitemShut {NoStop}%
\bibitem [{\citenamefont {Demetriou}\ \emph {et~al.}(2017)\citenamefont {Demetriou}, \citenamefont {Hewak}, \citenamefont {Ravagli}, \citenamefont {Craig},\ and\ \citenamefont {Kar}}]{demetriou2017nonlinear}%
  \BibitemOpen
  \bibfield  {author} {\bibinfo {author} {\bibfnamefont {G.}~\bibnamefont {Demetriou}}, \bibinfo {author} {\bibfnamefont {D.~W.}\ \bibnamefont {Hewak}}, \bibinfo {author} {\bibfnamefont {A.}~\bibnamefont {Ravagli}}, \bibinfo {author} {\bibfnamefont {C.}~\bibnamefont {Craig}},\ and\ \bibinfo {author} {\bibfnamefont {A.}~\bibnamefont {Kar}},\ }\href {https://doi.org/https://doi.org/10.1364/AO.56.005407} {\bibfield  {journal} {\bibinfo  {journal} {Appl. Opt.}\ }\textbf {\bibinfo {volume} {56}},\ \bibinfo {pages} {5407} (\bibinfo {year} {2017})}\BibitemShut {NoStop}%
\bibitem [{\citenamefont {Cohen}\ \emph {et~al.}(2003)\citenamefont {Cohen}, \citenamefont {Schwartz}, \citenamefont {Fleischer}, \citenamefont {Segev},\ and\ \citenamefont {Christodoulides}}]{cohen2003multiband}%
  \BibitemOpen
  \bibfield  {author} {\bibinfo {author} {\bibfnamefont {O.}~\bibnamefont {Cohen}}, \bibinfo {author} {\bibfnamefont {T.}~\bibnamefont {Schwartz}}, \bibinfo {author} {\bibfnamefont {J.~W.}\ \bibnamefont {Fleischer}}, \bibinfo {author} {\bibfnamefont {M.}~\bibnamefont {Segev}},\ and\ \bibinfo {author} {\bibfnamefont {D.~N.}\ \bibnamefont {Christodoulides}},\ }\href {https://doi.org/https://doi.org/10.1103/PhysRevLett.91.113901} {\bibfield  {journal} {\bibinfo  {journal} {Phys. Rev. Lett.}\ }\textbf {\bibinfo {volume} {91}},\ \bibinfo {pages} {113901} (\bibinfo {year} {2003})}\BibitemShut {NoStop}%
\bibitem [{\citenamefont {Eilbeck}\ \emph {et~al.}(1985)\citenamefont {Eilbeck}, \citenamefont {Lomdahl},\ and\ \citenamefont {Scott}}]{eilbeck1985discrete}%
  \BibitemOpen
  \bibfield  {author} {\bibinfo {author} {\bibfnamefont {J.~C.}\ \bibnamefont {Eilbeck}}, \bibinfo {author} {\bibfnamefont {P.}~\bibnamefont {Lomdahl}},\ and\ \bibinfo {author} {\bibfnamefont {A.~C.}\ \bibnamefont {Scott}},\ }\href {https://doi.org/https://doi.org/10.1016/0167-2789(85)90012-0} {\bibfield  {journal} {\bibinfo  {journal} {Physica D: Nonlinear Phenomena}\ }\textbf {\bibinfo {volume} {16}},\ \bibinfo {pages} {318} (\bibinfo {year} {1985})}\BibitemShut {NoStop}%
\bibitem [{\citenamefont {Richardson}\ \emph {et~al.}(2013)\citenamefont {Richardson}, \citenamefont {Fini},\ and\ \citenamefont {Nelson}}]{richardson2013space}%
  \BibitemOpen
  \bibfield  {author} {\bibinfo {author} {\bibfnamefont {D.~J.}\ \bibnamefont {Richardson}}, \bibinfo {author} {\bibfnamefont {J.~M.}\ \bibnamefont {Fini}},\ and\ \bibinfo {author} {\bibfnamefont {L.~E.}\ \bibnamefont {Nelson}},\ }\href {https://doi.org/https://doi.org/10.1038/nphoton.2013.94} {\bibfield  {journal} {\bibinfo  {journal} {Nat. Photonics.}\ }\textbf {\bibinfo {volume} {7}},\ \bibinfo {pages} {354} (\bibinfo {year} {2013})}\BibitemShut {NoStop}%
\bibitem [{\citenamefont {Miller}(2013)}]{miller2013reconfigurable}%
  \BibitemOpen
  \bibfield  {author} {\bibinfo {author} {\bibfnamefont {D.~A.}\ \bibnamefont {Miller}},\ }\href {https://doi.org/https://doi.org/10.1364/OE.21.020220} {\bibfield  {journal} {\bibinfo  {journal} {Opt. Express}\ }\textbf {\bibinfo {volume} {21}},\ \bibinfo {pages} {20220} (\bibinfo {year} {2013})}\BibitemShut {NoStop}%
\bibitem [{\citenamefont {Li}\ \emph {et~al.}(2013)\citenamefont {Li}, \citenamefont {Zhao},\ and\ \citenamefont {Vincent~Liu}}]{li2013topological}%
  \BibitemOpen
  \bibfield  {author} {\bibinfo {author} {\bibfnamefont {X.}~\bibnamefont {Li}}, \bibinfo {author} {\bibfnamefont {E.}~\bibnamefont {Zhao}},\ and\ \bibinfo {author} {\bibfnamefont {W.}~\bibnamefont {Vincent~Liu}},\ }\href {https://doi.org/https://doi.org/10.1038/ncomms2523} {\bibfield  {journal} {\bibinfo  {journal} {Nat. Commun.}\ }\textbf {\bibinfo {volume} {4}},\ \bibinfo {pages} {1523} (\bibinfo {year} {2013})}\BibitemShut {NoStop}%
\end{thebibliography}

%


\onecolumngrid

\vspace{1.5cm}

\newcommand{\beginsupplement}{%
        \setcounter{equation}{0}
        \renewcommand{\theequation}{S\arabic{equation}}%
        \setcounter{figure}{0}
        \renewcommand{\thefigure}{S\arabic{figure}}%
         }
 
 \beginsupplement

\section*{\large {Supplementary Information}\\ Nonlinear Switch and Spatial Lattice Solitons of Photonic ${\bf{ s}}$-${\bf{p}}$ Orbitals\\
\vspace*{0.2cm}}
\twocolumngrid

\section{Discrete Nonlinear Schr{\"o}dinger Equation}
\begin{figure}[]
\includegraphics[width=1\linewidth]{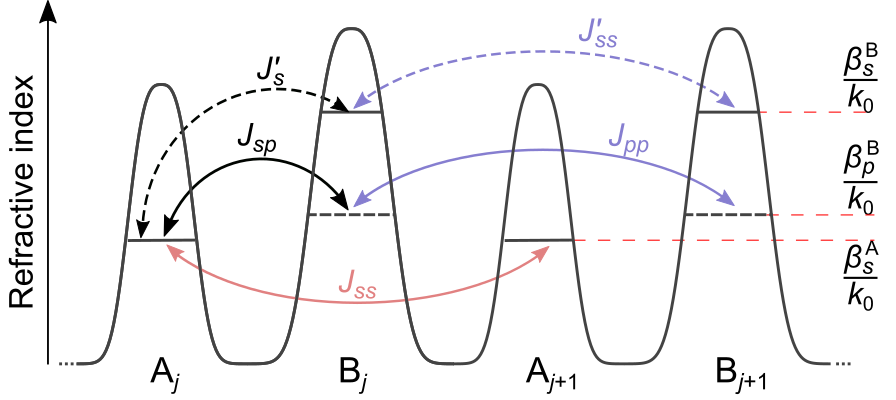}
    \caption{Similar to Fig.~\ref{Fig_model}(b) in the main text showing nearest and next-nearest neighbor couplings.}
    \label{supp_lattice}
\end{figure}

In the main text, we have presented the discrete nonlinear Schr{\"o}dinger equation~(\ref{DNLSE_sp1}, \ref{DNLSE_sp}) governing the light propagation in the $s$-$p$ array, ignoring the contribution from the $s$ orbital of B site. In this section, we further discuss the validity of such approximations. The interaction among all the three orbitals, i.e., $a_j^s$, $b_j^s$ and $b_j^p$, can be captured by the following equations
\begin{align}
 i\frac{\partial}{\partial z}a_j^s(z) \!=\! & -J_{sp} \big( b_j^p+ b_{j-1}^p \big) - J_{ss} \big( a_{j+1}^s + a_{j-1}^s \big) \nonumber\\
& - J'_s \big( b_j^s + b_{j-1}^s \big) -\beta_s^{\text{A}} a_{j}^s-g_s^{\text{A}}|a_j^s|^2 a_j^s  \, , \label{DNLSE_p_all1}\\
i\frac{\partial}{\partial z}b_j^p(z) \!=\! & -J_{sp} \big( a_j^s+ a_{j+1}^s \big)
-J_{pp} \big( b_{j+1}^p+b_{j-1}^p \big) \nonumber \\
&-\beta_p^{\text{B}} b_{j}^p-(g_p^{\text{B}}|b_j^p|^2+g_s^{\text{B}}|b_j^s|^2)b_j^p 
\, , \label{DNLSE_p_all2}\\
i\frac{\partial}{\partial z}b_j^s(z) \!=\! & -J'_{s} \big( a_j^s+ a_{j+1}^s \big)
-J'_{ss} \big( b_{j+1}^s+b_{j-1}^s \big) \nonumber \\
&-\beta_s^{\text{B}} b_{j}^s-(g_s^{\text{B}}|b_j^s|^2+g_p^{\text{B}}|b_j^p|^2)b_j^s \, ,
\label{DNLSE_p_all3}     
\end{align}
where $J'_s$ is the nearest neighbor coupling between the $s$ orbitals of A site and B sites. The next nearest neighbor coupling between the $s$ orbitals of B sites is given by $J'_{ss}$, Fig.~\ref{supp_lattice}. The significance of all other terms is given in the main text. In the limit of $|\beta_s^{\text{B}}-\beta_s^{\text{A}}|\! \gg \! J_{sp}$ and absence of optical loss, $\partial_z |b_j^s(z)|^2\!=\!0$.
In this situation, if the light is not coupled to the $s$ orbital of B site, Eqs.~(\ref{DNLSE_p_all1}-\ref{DNLSE_p_all3}) can be approximated by Eqs.~(\ref{DNLSE_sp1}, \ref{DNLSE_sp}). The above approximations are  valid for most results presented in the main text. However, for Fig.~\ref{Fig_P_soliton}, some amount of light is initially coupled in the $s$ orbital of B site. In this case, the nonlinear term in Eq.~\ref{DNLSE_sp} is modified as  $-(g_p^{\text{B}}|b_j^p|^2+g_s^{\text{B}}|b_j^s|^2)b_j^p$. Because of this, single-peak solitons at the $p$ orbital are experimentally observed at a relatively lower input power. If light is coupled only to the $p$ orbital of a B site, these solitons would be less localized at a given nonlinearity compared to those in Fig.~\ref{Fig_soliton}.

\section{More Details on Optical Characterizations}

\begin{figure}[t]
    \includegraphics[width=1\linewidth]{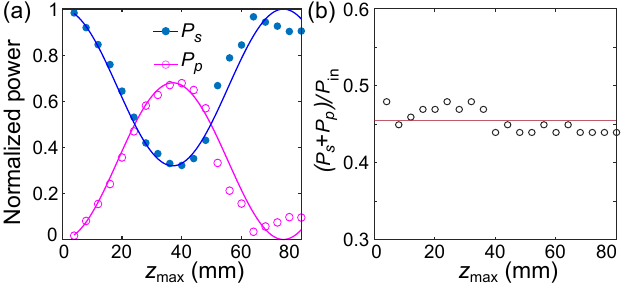}
    \caption{Linear characterization of the $s$-$p$ coupler as shown in Fig.~\ref{Fig_coupler_lin}(b) at $\lambda\!=\!1030$~nm wavelength of light. Here, $\Delta\!=\!0.048$~mm$^{-1}$ and $J_{sp}\!=\!0.035$~mm$^{-1}$. (b) Normalized transmitted optical power at the output of the $s$-$p$ coupler in (a) as a function of the interaction length $z_{\text{max}}$. The input power was $P_{\text{in}}\!=\!0.1$~mW for these measurements. 
 The red line indicates a mean transmission of $0.45$.}
    \label{Fig_coupler_lin_1030nm}
\end{figure}

{\it Loss measurements.$-$}
In this section, we discuss how the propagation loss of three different orbitals were measured or estimated.
It is straightforward to find the propagation loss of the $s$ orbital of A site. We first measure the insertion (propagation + input coupling) loss for two different lengths ($80$ and $40$~mm) of the single mode waveguide. Considering the cut-back method, the propagation loss  at $1030$~nm wavelength was found to be $0.25$~dB/cm from the difference in insertion losses.
For the $p$ orbital of the B site, measuring propagation loss is more challenging since a precise initial excitation of this orbital is required. To address this challenge, we use a set of $s$-$p$ couplers with varying length of the two-mode waveguide. We measure the insertion loss of the device by coupling light to the $s$ orbital of A site. Since the $s$ and $p$ orbitals of the neighboring site are near phase matched, light transfer occurs efficiently as shown in Fig.~\ref{Fig_coupler_lin_1030nm}(a). The $s$-$p$ couplers exhibited similar transmission (see Fig.~\ref{Fig_coupler_lin_1030nm}(b)) as that of an isolated single mode waveguide. For numerical simulations (e.g. Fig.~\ref{Fig_coupler_NL}(a)), the propagation loss for the $p$ orbital of B site $\alpha_p$ was considered to be $0.25$~dB/cm. 

\begin{figure}[h]
    \includegraphics[width=0.9\linewidth]{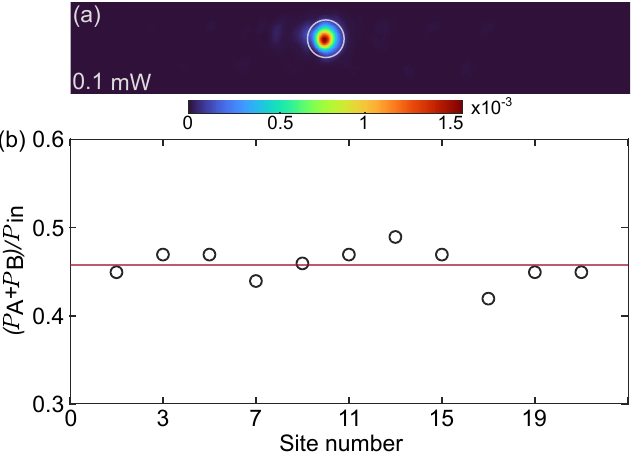}
    \caption{Linear characterization of the $s$-$p$ array shown in Fig.~\ref{Fig_model}(d) with input excitation at the $s$ orbitals of the B sites. Here $\lambda\!=\!1030$~nm 
    and $P_{\text{in}}\!=\!0.1$~mW.
    (a) Measured output intensity distribution with input coupling at the $s$ orbital of $11$-th B site (white circle). No significant tunneling of light was observed, confirming that $(\beta_s^{\text{B}}-\beta_s^{\text{A}})/J_{sp}\! \gg \!1$ and the value of $J'_{ss}$ is negligible for our experiments. The image is normalized such that the total power is $1$.  
    (b) Normalized transmitted output power for input coupling at the $s$ orbital of different B sites. The red line indicates a mean transmission of $0.46$.}
    \label{Fig_array_lin_1030nm}
\end{figure}

We then individually couple light at the center of each two-mode waveguide in the lattice. In this case, we can excite only the $s$ orbital of B site, and no measurable tunneling was observed after a propagation of $z\!=\!80$~mm, see Fig.~\ref{Fig_array_lin_1030nm}(a). We conclude that the $J'_{ss}$ is negligible in our experiment, and $\Delta'$ is significantly larger than $J_{sp}$. Additionally, the $s$ orbital of B site has a similar transmission (hence, the propagation loss) as that of an isolated single mode waveguide Fig.~\ref{Fig_array_lin_1030nm}(b).

To estimate the nonlinear refractive indices of the A and B sites, we study 
spectral broadening due to
self-phase modulation by coupling moderate-power laser pulses to the $s$ orbitals. We measure the broadening as a function of average input power and observed a linear variation for both waveguides. Assuming that the slope of the linear variation is proportional to the nonlinear strength of the orbital, we obtained $(n_2^\text{A}/n_2^\text{B})$.




\section{Self-consistency Algorithm}
\begin{figure}[t]
    \includegraphics[width=1\linewidth]{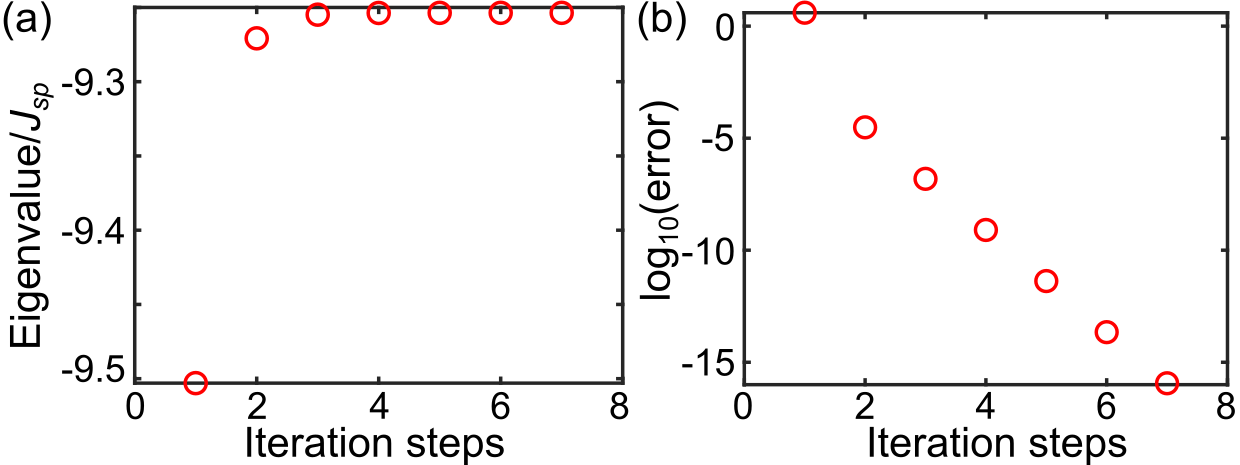}
    \caption{Self-consistency iteration algorithm. (a) Convergence of eigen value for a soliton peaked at $12^{\text{th}}$ site (Fig.\ref{Fig_NL_spectrum}(b)). (b) Calculated error is defined as $\epsilon\!=\!\sum |\psi^{n+1}-\psi^{n}|^2$ with each iteration step.}
    \label{Fig_soliton_convergence}
\end{figure}

We use self-consistency algorithm to iteratively calculate soliton solutions starting from a suitable initial guess state. Consider a finite array of $s$-$p$ orbitals with experimentally obtained parameters of the linear Hamiltonian $H_0$. For a chosen value of $g_s^{\text{A}}$, $g_p^{\text{B}}$ and power, the initial guess solution $\psi^n$ ($n$ indicates the iteration step) gives the nonlinearity-induced modification of on-site propagation constants and hence, the initial nonlinear Hamiltonian, $H_{\text{nl}}(z\!=\!0)$. The total Hamiltonian $H_t\!=\!H_0+H_{\text{nl}}$ is then diagonalized to obtain the eigenvectors and eigenvalues. We then find the eigenvector $\psi^{n+1}$ with maximal overlap with the normalized state $\psi^{n}$. In the next iteration step, 
$\psi^{n+1}$ is considered as the new initial state. The iteration process continues until the error, defined as $\epsilon=\sum |\psi^{n+1}-\psi^{n}|^2$ (the summation is over all sites) is sufficiently small, typically $10^{-10}$. Fig. \ref{Fig_soliton_convergence} shows the convergence of eigenvalue and the error in each iteration step in the case of the single-peak bulk soliton at $s$ orbital.

\begin{figure}[]
    \includegraphics[width=1\linewidth]{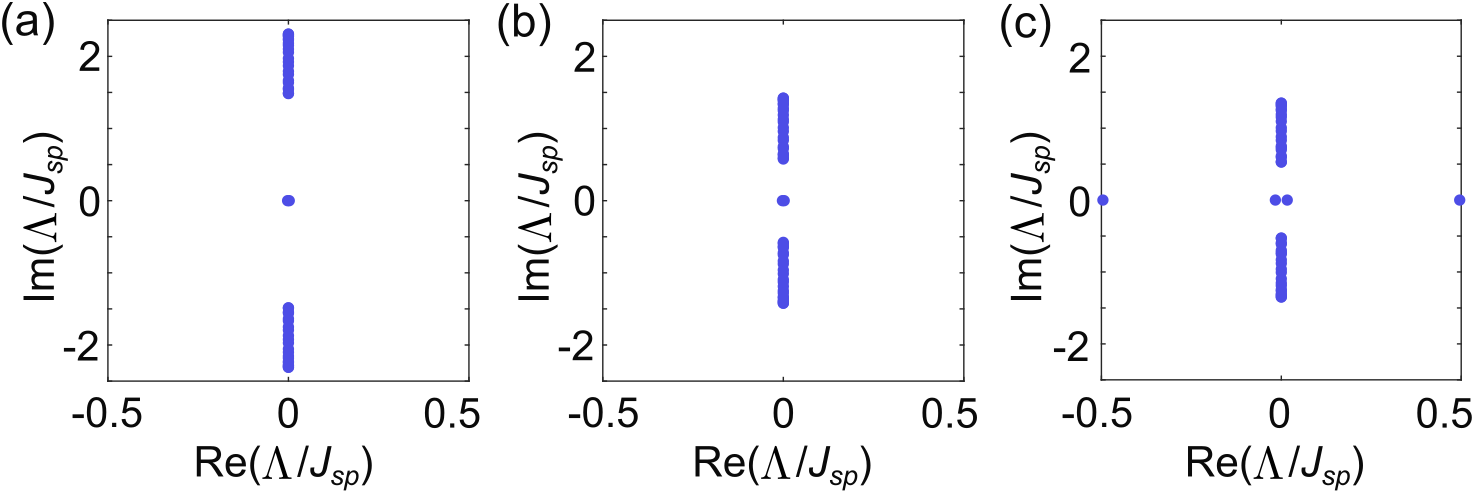}
    \caption{
    Stability analysis of the three types of solitons shown in Fig.~\ref{Fig_NL_spectrum}(b-d). The double peak soliton centered at the B site is linearly unstable, as indicated by the nonzero values of Re($\Lambda$).}
    \label{Fig_sol_stability}
\end{figure}

\section{Linear Stability Analysis}
In order to understand the stability of the soliton solutions obtained through self-consistency iteration algorithm, we perform the linear stability analysis~\cite{eilbeck1985discrete} 
for soliton solution of the form $\Psi\!=\!\psi \, \exp(-iE_0z)$, where $E_0$ is the eigenvalue. 
We analyze the stability of $\Psi$ by considering
a small perturbation
\begin{equation}
    \Psi\!=\!\big(\psi+ \epsilon (a+ib)\big) \exp(-iE_0z) \, ,
    \label{Soliton_noise}
\end{equation}
where $a$, $b$ are vectors of the form $a(b)\!=\!a'(b')\text{exp}({\Lambda z})$ and $\epsilon\ll1$. 
Solitons are linearly stable if the eigen values $\Lambda$ are purely imaginary. Fig.~\ref{Fig_sol_stability} shows the linear stability analysis for the three types of soliton solutions shown in Fig.~\ref{Fig_NL_spectrum}(b-d). For $P_{\text{in}}\!=\!10.5$~mW, the single peak solitons at A and B sites are linearly stable whereas the double-peaked soliton  centered at B site is linearly unstable having nonzero values of $\text{Re}(\Lambda)$.

\clearpage

\end{document}